\documentclass[journal=aanmf6,manuscript=regular]{achemso}

\usepackage[version=3]{mhchem} 
\usepackage{comment}
\usepackage{xr}
\makeatletter
\newcommand*{\addFileDependency}[1]{
  \typeout{(#1)}
  \@addtofilelist{#1}
  \IfFileExists{#1}{}{\typeout{No file #1.}}
}
\makeatother

\newcommand*{\myexternaldocument}[1]{%
    \externaldocument{#1}%
    \addFileDependency{#1.tex}%
    \addFileDependency{#1.aux}%
}
\myexternaldocument{text}

\setkeys{acs}{maxauthors = 0}

\SectionNumbersOff
\usepackage{booktabs}
\usepackage{graphicx}

\author{Anna Marzegalli}%
 
 \author{Francesco Montalenti}%
 
 \author{Emilio Scalise}
\email{emilio.scalise@unimib.it}
 
\affiliation{%
Department of Materials Science, University of Milano-Bicocca, Via Roberto Cozzi 55, 20125 Milan, IT
}%

\title[An \textsf{achemso} ]
  {Polytypic Quantum Wells in Si and Ge: \\ Impact of 2D Hexagonal Inclusions \\ on Electronic Band Structure}


\begin{document}





\begin{abstract}
Crystal defects, traditionally viewed as detrimental, are now being explored for quantum technology applications. This study focuses on stacking faults in silicon and germanium, forming hexagonal inclusions within the cubic crystal and creating quantum wells that modify electronic properties. By modeling defective structures with varying hexagonal layer counts, we calculated formation energies and electronic band structures. Our results show that hexagonal inclusions in Si and Ge exhibit a direct band gap, changing with inclusion thickness, effectively functioning as quantum wells. We find that Ge inclusions have a direct band gap and form Type-I quantum wells. This research highlights the potential of manipulating extended defects to engineer the optoelectronic properties of Si and Ge, offering new pathways for advanced electronic and photonic device applications.\end{abstract}

\section{Introduction}
Crystal defects are distortions of the periodic crystal lattice that are typically undesirable because they deteriorate the mechanical, electrical, and optical properties of solids. Despite this, solid-state point defects that act as single-photon sources and quantum bits are among the most promising candidates for quantum technologies, with a long history in solid-state physics and quantum information science\cite{Wolfowicz2021}. In contrast, extended defects are often considered unwanted and detrimental for opto- and power-electronic devices\cite{Kato2022,Iwata2003,Scalise2020,Perera,Joyce}. However, these defects are intriguing from a physical perspective and hold potential for innovative applications. They can form quantum wells within the material and, if repeated periodically along the stacking direction, they may give rise to superlattices\cite{Bechstedt1995}. The most common and straightforward type of extended defects, stacking faults (SFs), occur quite naturally in many materials that exhibit polytypism. SFs can be viewed as inclusions of a few layers of one polytype within the perfect layer stacking of another polytype\cite{Scalise2019}. They are generally not considered the most significant defects in comparison to other types of defects in silicon (Si) and germanium (Ge), such as dislocations, vacancies, and interstitials. Nonetheless, errors in the stacking sequence are often observed in Si and Ge nanostructures and particularly in nanowires \cite{He2019,Heine1992,Tang2017,Amato2017}. Moreover, recent experiments report on hexagonal germanium (hex-Ge) nanostructures within epitaxially grown cubic-Ge on silicon (Si(001)) substrates. These hexagonal inclusions, which are essentially stacking faults, form as a result of strain-induced nanoscale crystal structure transformations under far-from-equilibrium growth conditions. In the same study, Zhang and coauthors inferred that these hexagonal-Ge nanostructures have direct-band gap features\cite{Zhang}.

In this work, we perform first-principles calculations to determine the electronic band structures of these hexagonal inclusions in Si and Ge. Our results prove that Ge inclusions exhibit a direct band gap, with the band gap energy modulated by the thickness of the inclusions.

\section{Results and discussion}

\begin{figure}
  \includegraphics[width=0.95\textwidth]{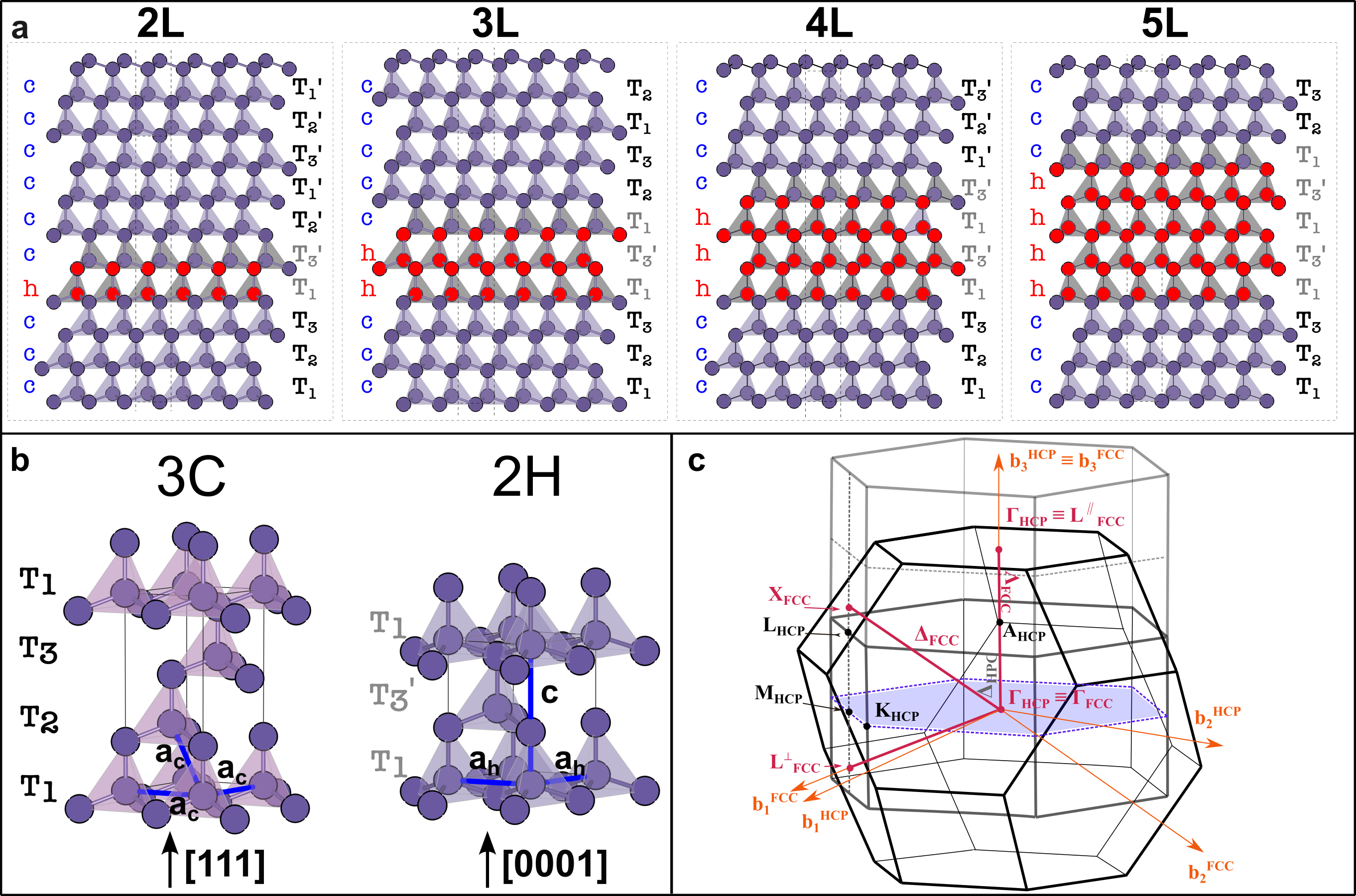}
 \caption{a) Tetrahedral stacking sequence of the different defective structures modelled. The red and plum triangles highlight the twinned or normal tetrahedra and correspond to down or up spin configurations of the layers, according to the axial next-nearest-neighbor Ising (ANNNI) model  \cite{ANNNI}. The \textit{h} and \textit{c} letters indicate hexagonally and cubically stacked layers, respectively. Similarly, atoms having all of their first and second nearest neighbours positioned on hexagonal (cubic) diamond lattice sites are coloured in red (plum).  $T_{1,2,3}$ and $T_{1,2,3}'$ labels indicate the tetrahedral stacking sequences of the different structures. b)  Tetrahedral stacking sequence of the cubic and hexagonal diamond crystals together with their primitive vectors $a_c$, $a_h$ and $c$. Both are represented using hexagonal cells made of three and two layers, for the cubic and hexagonal one, respectively. c) FCC and HCP Brillouin zones are illustrated in black and grey colours, respectively. Their $\Gamma$ points are superimposed with the $b_3$ reciprocal vectors oriented in the same direction. Because the volume of the HCP BZ is half of the FCC one, a periodic replica of the HCP Brillouin zone is also shown in light grey. Note that the drawing would change slightly if the (hexagonal) reciprocal lattice of the 3C crystal in the hexagonal cell, as shown in b), is represented instead of the HCP Brillouin zone: the hexagonal lattice shown in gray in c) has shrunk along $b_3$. Accordingly, one gets $M_{HCP} \equiv L^{\perp}_{FCC}$ and $L_{HCP} \equiv X_{FCC}$, but $\Gamma_{HCP} \not\equiv L^{\parallel}_{FCC}$. This means that when using a hexagonal cell instead of the 3C primitive one, the $L$ point will be folded to $X$ (not to $\Gamma$, see Fig. \ref{fgrband} in the ESM).}
 \label{fgrstructure}
\end{figure}

SFs in the cubic diamond crystal alter its typical stacking sequence of tetrahedra ($T_1T_2T_3$) along the $[111]$ direction. The defective cubic crystal becomes non-periodic along the latter direction, and new sequences ascribable to the 2H tetrahedral stacking sequence (e.g. $T_1T_3'$) appear in between the 3C stacking sequences. These planar defects constitute 2D-hexagonal inclusions in the 3C crystal structure, as evidenced in Fig. \ref{fgrstructure}a. We modelled defective structures that include between 2 and 5 layers (named 2L-5L), which can be identified with the hexagonal stacking sequence. In other words, we include from one to four 2H unit cells in a large supercell of 3C crystal, which is ideally semi-infinite along the $[111]$ direction. The 2L and 3L structures are better known in the literature as twin boundaries (TB) and intrinsic stacking fault (ISF), respectively \cite{Keller2023,Mattheiss1981,Chou1985,denteneer1988,Oleg2019}. We report in Table \ref{TABI} the formation energies calculated as the energy difference (per atom) between all defective structures shown in Fig. \ref{fgrstructure} and equivalent pristine 3C cells. For both Si and Ge, these values are very similar to those obtained using the ANNNI model and exploiting the polytype cohesive energies, as described in the method sections. Although no previous calculations were found for the 4L and 5L structures, the formation energies calculated for the TB and ISF (i.e., 2L and 3L structures, respectively) agree well with the values reported most recently in the literature \cite{Keller2023}. These energies are calculated within the framework of density functional theory (DFT) using the generalized gradient approximation (GGA) and a modified version of the Perdew-Burke-Ernzerhof (PBE) functional optimized for solids \cite{PBEsol}. For the formation energy values calculated with the ANNNI model, we also include values obtained by exploiting the non-empirical strongly constrained and appropriately normed (SCAN) meta-generalized gradient approximation (meta-GGA) exchange-correlation functional \cite{SCAN,SCAN2}, which has been shown to provide more accurate energies of several molecules and materials\cite{Sun2016}. Indeed, especially for Si the calculated value with SCAN for the ISF is closer to the experimental value of $55\pm7$ $mJ/m^{2}$ \cite{SFenergies}. \\
\begin{table}
    \centering
\begin{tabular}{lcccccccl}\toprule[1.5pt]\midrule
 \multicolumn{4}{c}{Supercell} & \multicolumn{4}{c}{ANNNI}
\\\cmidrule(lr){2-5}\cmidrule(lr){6-9}
                &$\gamma_{2L}$&$\gamma_{3L}$&$\gamma_{4L}$&$\gamma_{5L}$&$\gamma_{2L}$&$\gamma_{3L}$&$\gamma_{4L}$&$\gamma_{5L}$\\\cmidrule[1.5pt](lr){1-5}\cmidrule[1.5pt](lr){6-9}

            Si & 9.28 & 34.59 & 60.06 & 85.81 & 7.66 & 35.17 & 55.32 & 82.84 \\
              &   &   &   &   & \textbf{19.29} & \textbf{51.13} & \textbf{78.61} & \textbf{110.46} \\
            Ge & 31.02 & 70.20 & 109.74 & 146.22 & 31.30 & 69.51 & 105.49 & 143.70 \\
            &   &   &   &   & \textbf{42.51} & \textbf{81.08} & \textbf{119.30} & \textbf{157.87} \\\bottomrule[1.5pt]
\end{tabular}

 \caption{Formation energies $\gamma$ (in $mJ/m^2$) calculated using PBEsol functionals and exploiting the energy of defective supercells or the ANNNI model. The bold text highlights the values obtained by using the SCAN exchange-correlation functionals.}
 \label{TABI}     
\end{table}

One may notice that the formation energies in Ge are about twice as large as the corresponding ones in Si. However, especially for Si, these values are not much higher than the formation energies calculated for the ISF in other materials with the zinc-blende structure such as III-Vs or II-VIs  \cite{SFenergies,GlasSF} or even SiC \cite{Scalise2019}, where this kind of defects are very often experimentally observed. This suggests that to get inclusions of the 2D hexagonal layers in Si and Ge, synthesis conditions and kinetic aspects are more important than thermodynamics. Thus, special expedients such as exploiting strain and far-from-equilibrium growth conditions as in the recent work of Zang and coauthors \cite{Zhang}, may allow one to obtain and engineer 2D hexagonal inclusions in 3C crystals. Then, understanding the optoelectronic properties of these Si and Ge structures is crucial. We provide some fundamental knowledge by the electronic band structures calculated \textit{ab initio} and discussed below.\\
3C-Si and Ge have a crystal structure with a face-centred cubic (FCC) Bravais lattice and 2 atoms in the basis. The corresponding first Brillouin zone (BZ) is illustrated in Fig. \ref{fgrstructure}c with black lines. The hexagonal phases (2H) have a hexagonal closed packed (HCP) lattice with 4 atoms per unit cell, and its corresponding BZ is also illustrated in Fig. \ref{fgrstructure}c with dark grey lines. Both structures have tetrahedral coordination but with a different stacking sequence of the tetrahedron in the $[111]$ and $[0001]$ directions for 3C and 2H, respectively (see Fig. \ref{fgrstructure}b). Considering an ideal hexagonal lattice with $c/a=\sqrt{8/3}$ and identical in-plane lattice parameters as the cubic lattice ($a_c=a_h$ in Fig. \ref{fgrstructure}b), the cubic primitive cell will have half of the volume compared to the hexagonal one. Then, the folding of the cubic diamond band structure in the hexagonal Brillouin zone (dark grey lines in Fig. \ref{fgrstructure}c) is expected. In particular, the $L_{FCC}$ symmetry point becomes coincident with the $\Gamma_{HCP}$ point of the replica of the central cell. Thus, the bands at $L_{FCC}$ backfold to $\Gamma_{HCP}$, and for Ge this induces the $L_{6}$ conduction band minimum of the 3C to backfold to the $\Gamma_{8}$ conduction band minimum at the zone center of the 2H, explaining the reason for the direct gap in 2H-Ge \cite{De_2014, Rodl2019}. 

\begin{figure}
  \includegraphics[width=0.95\textwidth]{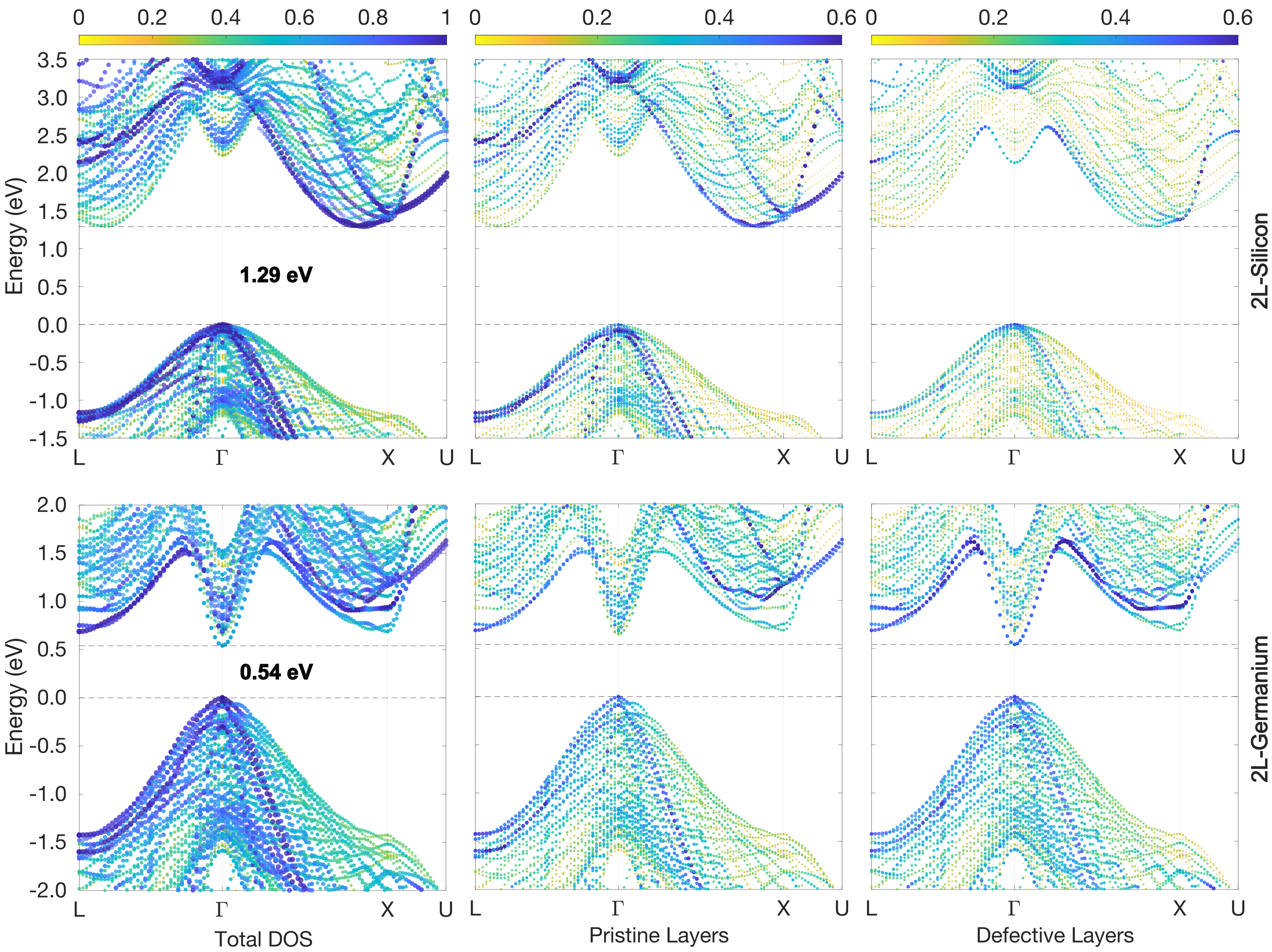}
  \caption{2L-Si and -Ge band structures (top and bottom panels, respectively) unfolded into the primitive cubic cell (colour scale). Energies are plotted relative to the top of the VB. The middle and right panels highlight the different contributions to the band states of the pristine-like and defective atoms (red and plum atoms, respectively, as shown in Fig. \ref{fgrstructure}). This is obtained by further weighting the colour intensity by the normalized local density of states of the different atoms. }
  \label{fgr:2L}
\end{figure}

When 2D hexagonal layers are inserted into a perfect 3C-Si or Ge crystal, the cubic crystal periodicity along the $[111]$ direction is broken and hexagonal supercells are more convenient to model the defective crystals. In principle, the pristine 3C crystal could also be modelled using a hexagonal supercell with three layers having the ($T_1T_2T_3$) stacking sequence, as illustrated in Fig. \ref{fgrstructure}b. The hexagonal unit cell has the same in-plane lattice parameters ($a_c$) as the primitive cubic cell but with a vertical size equal to $3\times a_c\sqrt{2/3}$, thus being $3/2$ the c parameter of the ideal 2H cell. If the 3C structure is modelled with such a hexagonal cell, band unfolding is necessary to correctly visualize the band structure along the high-symmetry points of the FCC Brillouin zone. Firstly, we validated this approach by calculating the band structure for both 3C-Si and Ge in a hexagonal cell containing 18 layers, thus replicating 6 times the hexagonal unit cell in the $[111]$ direction, and shown in Fig. \ref{fgrband} of the ESM. In the Si case, the folding of the bands does not modify the nature of the gap, but the minimum of the conduction band (CB) near the X point is backfolded near the L point, this muddles the order of the band transitions. The situation is even worse in Ge, where the band folding creates three minima of the CB at the $L$, $\Gamma$ and $X$ points with roughly the same energy. We verified that after the band unfolding, the expected band structures for both Si and Ge are obtained, showing a gap of about 1.27 eV for Si and 0.66 eV for Ge. These results are obtained by using the modified Becke-Johnson (MBJ) exchange functional, which has been shown to provide very accurate band gaps both for 3C and 2H-Si and Ge, similar to the hybrid HSE06 functional but with considerably less computational effort \cite{Rodl2019,Keller2023,Tran_2007,TB09,META,BJ06}.\\
Although the 2H inclusions break the cubic symmetry and also slightly modify the atomic positions of the 3C-like layers, the band unfolding is still very effective in identifying backfolded bands, which appear in the band structure plots of such defective structures. Fig. \ref{fgr:2L} illustrates the calculated band structure for the 2L structures of Si and Ge for which we used a colour scale to dim the backfolded bands from the rest. One may indeed notice (Fig. \ref{fgr:2L}) that the more intense points, in dark blue, form bands closely resembling the bulk Si and Ge band structures (unfolded band structures plotted in Fig. \ref{fgrband} of the ESM can be used for the comparison).  But the minimum of the CB at the $\Gamma$ point, which appears very intense (blue colour) both in Si and Ge is strange, in light of the above discussion on the band folding: bands backfolded at the $\Gamma$ point, due to the repetition of the 3C cell along the $[111]$ direction, should almost disappear after band unfolding. Focusing on the Si case, in the top-left panel of Fig. \ref{fgr:2L} and making the comparison with the pristine Si band structure (shown in Fig. \ref{fgrband} of the ESM), the minimum of the CB at $\Gamma$ would be expected at around 3.25 eV from the valence band (VB) edge. Dark blue bands are indeed present in the band structure of the 2L-Si structure, at about 3.25 eV and around the $\Gamma$ point, but other branches at lower energies also appear in blue colour.  Weighting additionally the colour intensity by the local density of states of atoms belonging to the 3C-like layers or 2H inclusions (plum and red atoms, respectively, in Fig. \ref{fgrstructure} ), we can distinguish clearer the contributions of the hexagonal inclusions to the band structure, as illustrated in the middle-right panels of Fig. \ref{fgr:2L}. The CB minimum at about 2.3 eV in  $\Gamma$ is indeed attributable mainly to the two hexagonal layers. Being the hexagonal inclusions non-periodic along the $[111]$ direction (or $[0001]$ for the hexagonal crystal), they have a 2D-hexagonal BZ, which is illustrated by the blue plane in Fig. \ref{fgrstructure}c. Thus, in this case, the backfolding in $\Gamma$ of the energy states in the $[111]$ direction is an actual physical effect but not merely an artifact of the supercell. 
This effect is even more peculiar in Ge, because similarly to the bulk 2H case \cite{De_2014}, the bottom of the CB appears at the $\Gamma$ point, thus leading to the direct bandgap. This is due to the zone folding and the hexagonal crystal field: the latter is responsible for splitting the degenerate top valence band levels at $\Gamma$, forming two bands, the fourfold-degenerate $\Gamma_{6}$ and twofold $\Gamma_{1}$, which are separated by the hexagonal crystal field energy when spin-orbit interactions are still neglected. When also the spin-orbit interactions are considered, the $\Gamma_{6}$ splits into the $\Gamma_{9}$ heavy hole and $\Gamma_{7}$ light hole state, with the $\Gamma_{1}$ getting the $\Gamma_{7}$ crystal-field split-off state \cite{De2010} (see Fig. \ref{fgrband} in the ESM). In other words, the large crystal field energy in hexagonal Ge up-shifts the top of the VB level, and is thus responsible for the reduction of the direct energy gap at $\Gamma$ in 2H-Ge, as compared to the indirect 3C band-gap \cite{Rodl2019,De_2014}. 
Indeed, a built-in crystal field due to the 2D-hexagonal inclusions into the 3C-phase can be assumed for the defective structure considered here. In turn, this crystal field is expected to reduce the energy gap at $\Gamma$ as compared to the pristine case. This is indeed observed in Fig. \ref{fgr:2L} for the Ge case, and the right panel attributes the bottom of the CB at the zone center to the hexagonal layers, with a reduced gap of about 0.54 eV. On the contrary,  the central panel of Fig. \ref{fgr:2L}, which highlights only the contribution of the pristine layers to the band structure, confirms that it is closer to the band structure of the bulk 3C-Ge, thus with an indirect gap.
\begin{figure}
  \centering
    \includegraphics[width=0.95\textwidth]{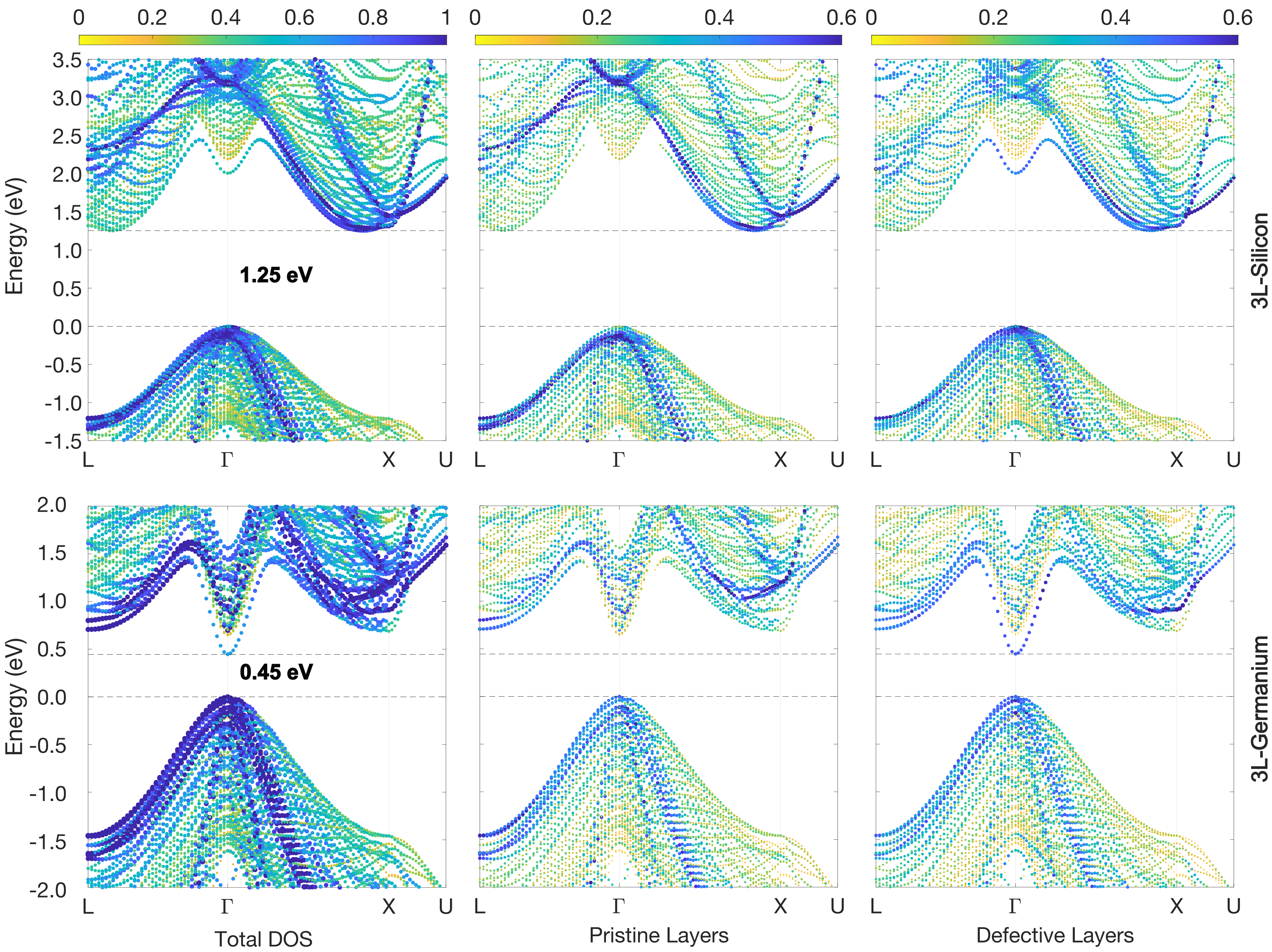}
  \caption{3L-Si and -Ge band structures (top and bottom panels, respectively) unfolded into the primitive cubic cell (colour scale). Energies are plotted relative to the top of the VB. The middle and right panels highlight the different contributions to the band states of the pristine-like and defective atoms (red and plum atoms, respectively, as shown in Fig. \ref{fgrstructure}). } \label{fig:3L}
\end{figure}

\begin{figure}
  \centering
  \includegraphics[width=0.95\textwidth]{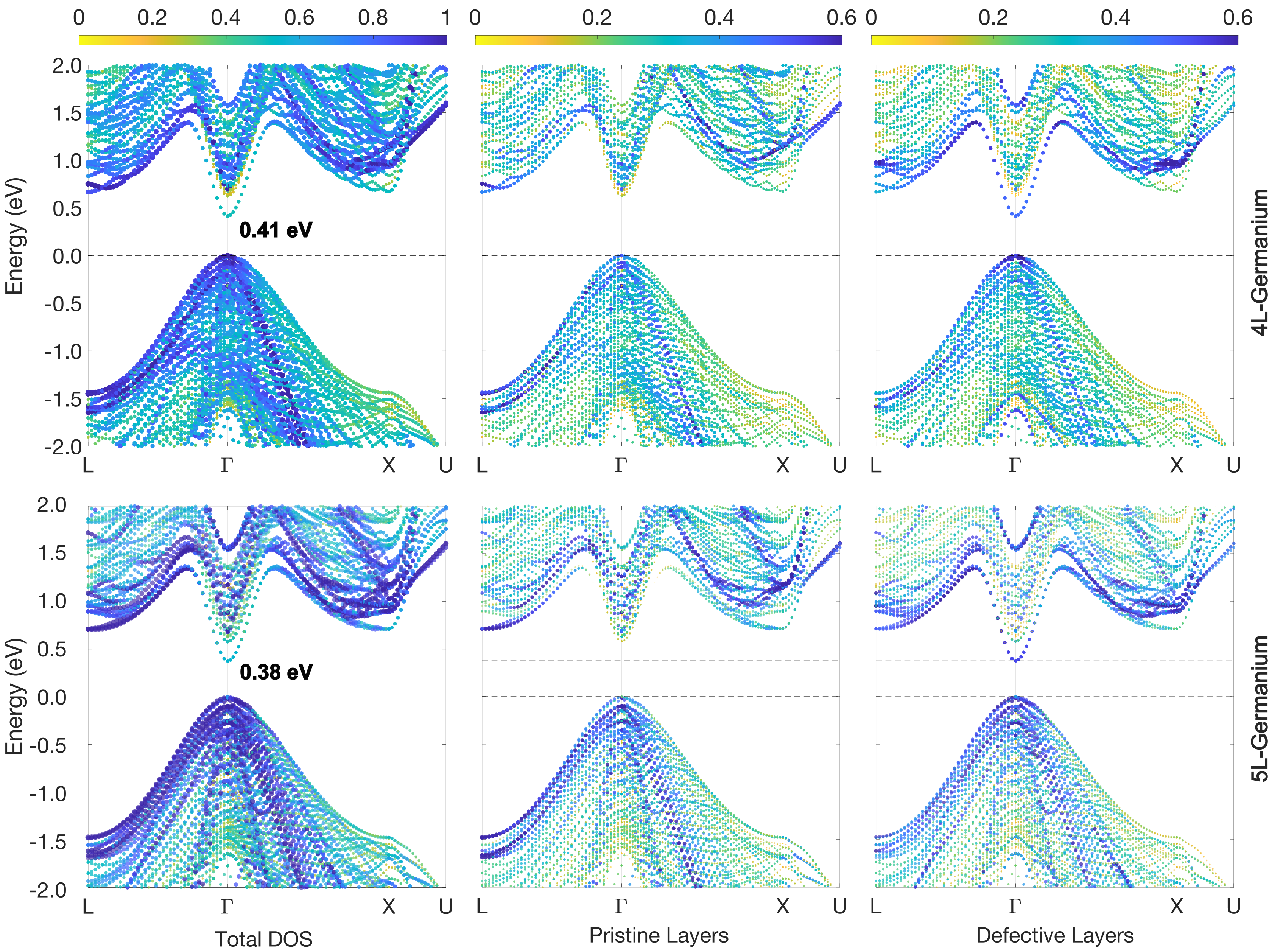} 
  \caption{4L-Ge and 5L-Ge band structures (top and bottom panels, respectively) unfolded into the primitive cubic cell (colour scale). Energies are plotted relative to the top of the VB. The middle and right panels highlight the different contributions to the band states of the pristine-like and defective atoms (red and plum atoms, respectively, as shown in Fig. \ref{fgrstructure}).
  } 
  \label{fig:45L}
\end{figure}

The energy gap reduces further in the case of the 3L structures, with values of 1.25 eV and 0.45 eV for Si and Ge, respectively, as illustrated in Fig. \ref{fig:3L}. This is not unexpected considering that the crystal field energy has been shown to increase linearly with the percentage of hexagonality of the Si and Ge crystal phases \cite{Keller2023}, which is defined as the ratio between the number of hexagonal and cubic layers. In principle, the defective structure shown in Fig. \ref{fgrstructure}a would be composed of an infinite number of cubic layers with a few inclusions of hexagonal layers. However, if we consider just a fixed number of total layers around the inclusions, beyond which the effect of the hexagonal inclusion vanishes, then one could assess that the hexagonality of the defective structures evidently increases with the number of hexagonal inclusions. Hence, the crystal filled would also increase, and in turn, a decrease in the gap is expected when the number of hexagonal inclusions increases. This tendency is confirmed by the results of the 4L and 5L structures, whose band structures for Ge are illustrated in Fig. \ref{fig:45L}, while the corresponding graphs for Si are included in Figs. \ref{fgr:4L-Silicon} and \ref{fgr:5L-Silicon} of the ESM. 

Note that the decrease in the band gap is much less evident in Si compared to Ge, because in the former case, the bottom of the CB is not at the $L$ point as for Ge, thus the $L_{FCC} \rightarrow \Gamma_{HCP}$ band folding does not change the nature of the gap in this case. Still, the direct gap at $\Gamma$ is also reduced for Si with the 2D hexagonal inclusions but remains much larger than the indirect band gap. The latter is practically identical for 2L compared to the calculated 3C-Si value, but slightly decreases with the number of hexagonal inclusions, reaching the minimum of 1.21 eV for the 5L-Si case (shown in Fig. \ref{fgr:5L-Silicon} of the ESM). This trend is again in qualitative agreement with the decrease in the indirect band gap energy with hexagonality, as predicted for Si polytypes \cite{Keller2023}. Although the indirect band gap of 2L-Si is very close to that of 3C-Si, the spatially resolved band structures in Fig. \ref{fgr:2L} reveal small changes in the band edges between the defective and the pristine part of the structure. In particular, the top of the valence band at $\Gamma$ rises by about 80 meV moving from the pristine 3C layers to the hexagonal inclusions, while the bottom of the conduction band (close to $X$) is almost not shifting at all. This means that the thinnest inclusion of 2D-hexagonal layers in 3C-Si (2L structure) forms a quasi-Type-II quantum well, with holes trapped in the well and electrons more delocalized in all the layers (predicted band offsets for the CB and VB are $\Delta E_C=0$ and $\Delta E_V=80meV$, respectively). The visualization in real space of the wavefunctions corresponding to the band edges of 2L-Si confirms this conclusion and is shown in the ESM (Fig. \ref{fgr:2L-Silicon-charge}). When the number of hexagonal layer inclusions in Si increases, also the bottom of the conduction band rises further, and Type-II quantum wells are formed with localization of holes in the 2D-hexagonal layers and electrons in the pristine 3C layers, as confirmed by visualization in real space of the wavefunctions corresponding to the band edges of 5L-Si (shown in Fig. \ref{fgr:2L-Silicon-charge} of the ESM). The largest band offsets are observed for the 5L structure, with values of $\Delta E_C=-40$ and $\Delta E_V=120meV$.

\begin{figure}
  \includegraphics[width=0.95\textwidth]{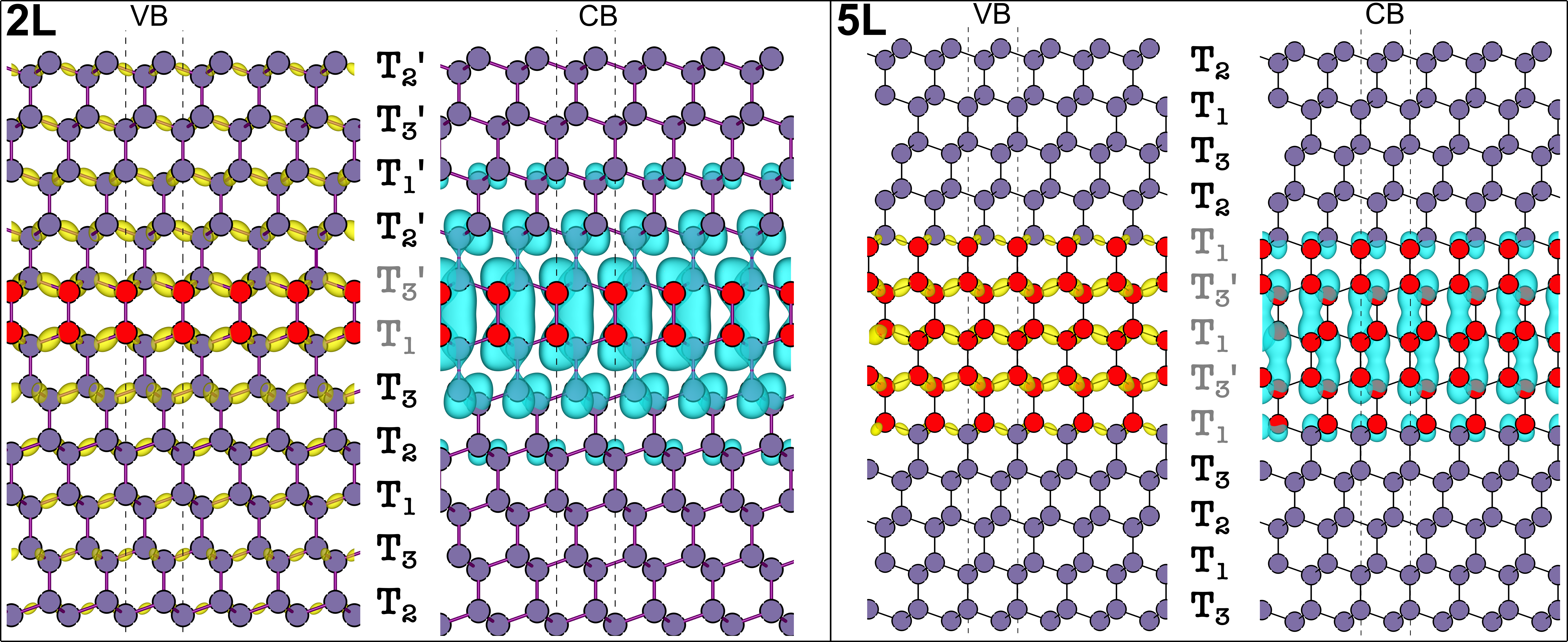}
  \caption{Real-space wavefunction (squared) for the states at the VB and CB edges of the 2L and 5L-Ge. }
  \label{fgr:charge}
\end{figure}

In Fig. \ref{fgr:charge} the visualization in real space of the wavefunctions \cite{vaspkit, WAVETRANS} for the Ge structures is shown, particularly for the 2L and 5L structures.  In these cases, because of the $L_{FCC} \rightarrow \Gamma_{HCP}$ band folding and the consequent bottom of the CB at $\Gamma$ for the 2D hexagonal Ge inclusions, the CB offsets are inverted compared to the Si case. In fact, the wave functions corresponding to the states at the CB edge are localized within the 2D hexagonal inclusions. Thus, in Ge, the quantum wells formed by the hexagonal inclusions trap electrons in the well. However, in the 2L-Ge case, the top of the VB for the 2D inclusions and the pristine layers are quite aligned (see Fig. \ref{fgr:2L}), and in fact the wavefunctions corresponding to VB edge states are quite delocalized in the whole structure, as illustrated in the left panels of Fig. \ref{fgr:charge}, providing again a quasi Type-II quantum well ($\Delta E_V=0$ and $\Delta E_C=100meV$). Similarly, the 3L-Ge remains a quasi-Type-II quantum well ($\Delta E_V=0$ and $\Delta E_C=180meV$). However, by increasing the number of the 2D hexagonal Ge inclusions, the valence band offsets increase and Type-I quantum wells are formed in the case of the 4L and 5L structures ($\Delta E_V=12$ and $15 meV$, $\Delta E_C=210$ and $250 meV$, for4L and 5L, respectively), the latter is shown with the plot of the wavefunctions in real space evidencing the trapping of both charge carriers in the quantum well. \\
In general, our calculated band offsets are qualitatively in agreement with the band alignment calculated for the polytypic Si and Ge homojunction \cite{Keller2023} and also with the predicted band offsets of the 3C / 2H junctions of Si and Ge \cite{Amato2017}. However, we show that the inclusions of 2D hexagonal layers in Si and Ge form very thin quantum wells, with a width between $\sim6.3\AA$ and $\sim15.7\AA$ for Si or $\sim6.5\AA$ and $\sim16.5\AA$ for Ge. They are Type-II quantum wells with indirect band gaps for Si and Type-I quantum wells with direct band gaps in Ge, but both get quasi-Type-II for the thinner inclusions. We show that the band offsets of these quantum wells are mainly driven by the changes in the stacking sequence. Particularly for Ge, these quantum wells can take advantage of the reduced dimensionality, and especially of the Brillouin zone-folding conferring the direct bandgap,  thus leading to an enhancement of optical transitions. Although not explicitly shown here for the 2D-hexagonal inclusions, another typical advantage of structures with reduced dimensionality is that quantum confinement enhances the oscillator strength \cite{Barbagiovanni2014}. Moreover, local asymmetry caused by random insertions and different distances between the 2D hexagonal structures can cause breaking of the \textbf{k}-conservation rules and enhance further the optical transitions, as in the case of hexagonal silicon-germanium alloys \cite{Belabbes2022, Borlido2021}.  Our results and the discussion above would indeed support the strong photoluminescence (PL) peak observed recently by Zhang \textit{at al.} \cite{Zhang}. However, the agreement is qualitative and reinforces the general tendency of the 2D hexagonal inclusions in Ge to enhance the optical transitions. Still, our predictions of the bandgap for the 2D hexagonal inclusions ($0.38eV<E_g<0.54eV$) are quite smaller than the exact value derived from the experimental observed PL peak ($\sim0.79 eV$) \cite{Zhang}.  Moreover, by including different numbers of hexagonal inclusions as in these recent experiments (at least 2L, 3L and 4L structures were observed together) one would expect a few different PL peaks because the predicted direct bandgaps are different, and the different quantization of energy levels is expected to form also additional subbands, due to the different width of the hexagonal inclusions. This latter scenario agrees with the theoretical and experimental findings for Wurtzite inclusions in Zincblende GaP \cite{Gupta2019}, but does not agree well with the recent observations for the Ge case\cite{Zhang}.

\section{Conclusion}
In conclusion, our study reveals that hexagonal inclusions within cubic Si and Ge structures can induce significant electronic band structure modifications by forming diverse types of quantum wells. Specifically, the inclusion of 2D hexagonal layers reduces the band gap, leading to a direct band gap in Ge. This finding opens new avenues for tailoring the optoelectronic properties of Si and Ge through the controlled introduction of stacking faults and hexagonal inclusions, with potential applications in quantum technology and photonics.

\section{Method}
\textit{ab initio} calculations were performed within the Density Functional Theory and the generalized gradient approximation (GGA)~\cite{Perdew1996} and a modified version of the Perdew-Burke-Ernzerhof (PBE) functional optimized for solids \cite{PBEsol}. Projected augmented wave (PAW) pseudopotentials \cite{Blochl1994,Kresse1999} were used, as implemented in the VASP code \cite{Kresse1996}. The energy cut-off was set to 500 eV and a (16×16×8) k-point mesh was used to minimize the total energy of the hexagonal unit cell structure used to model the 3C-Si and Ge, reducing the mesh for the larger slab calculations according to the ratio of their size and the unit cell. All structures were relaxed until the forces on all atoms were less than $10~\text{meV}/\text{\AA}$.
Larger supercells repeated in the $[111]$ direction have been used to model the defective structure, with an in-plane lattice parameter equal to that of the 3C structure. They included 48, 33, 30 and 39 atomic layers for the 2L, 3L, 4L and 5L structures, respectively.  For the 2L and 4L structures, two different hexagonal inclusions equally spaced by 3C layers were needed to keep the periodic boundary conditions along the $[111]$ direction. Three different hexagonal inclusions were needed for the 3L and 5L structures on the other ends, to keep an integer proportionality factor between the lattice size in the $[111]$ direction of the supercell and that of the hexagonal unit cell of the 3C structures, allowing the band unfolding. The fold2Bloch code \cite{Unfold, fold2bloch} was used to unfold all band structures into the primitive cubic cell. The local density of state was then used to further weight the band structure by the contribution of the defective part (hexagonal inclusions) and pristine part (3C crystal). The vaspkit tool \cite{vaspkit,WAVETRANS} was used for the calculation of the real-space squared wave functions. All the atomistic structures are drawn using the VESTA software \cite{VESTA}. 
 
Electronic band structures were obtained using the modified Becke-Johnson (MBJ) exchange correlation functional, which has been shown to provide very accurate band gaps both for 3C and 2H-Si and Ge, similar to the hybrid HSE06 functional but with considerably less computational effort \cite{Rodl2019,Keller2023,Tran_2007,TB09,META,BJ06}. In the latter case, a reduced (4×4×1) k-point mesh was used to obtain a converged wavefunction that was then used for the MBJ band structure calculation, using 120 k-points along high symmetry points $L - \Gamma - X -U$.

\subsection{Axial next neighbour Ising model (ANNNI)}

Following the description proposed by Heine \cite{Heine1992} the tetrahedra composing 3C and 2H structures (labelled by $T$ and $T'$ in the main section) can be associated with a spin value ($\sigma$=+1 or -1) in dependence of their orientation. The tetrahedra $T$ in the cubic sequence are defined as positive in spin. Consequently, the energy for each bilayer in structures composed of n tetrahedra (bilayers) at T = 0K can be written as:

\begin{equation}
    E =E_0 - \frac{1}{n}\sum_{i=1}^{n}\sum_{j=1}^{\infty}J_j\sigma_i\sigma_{i+j}
\end{equation}

Where $E_0$ is independent energy taken as a reference, the label $i$ counts the bilayers in the unit cell of the structures, and $j$ runs over all the interacting bilayers. The $J_j$ parameters are the interaction energies between the considered layer and the $j$th-neighbour bilayers. Interaction energies between bilayers over the third order are neglected becoming usually very small. $J_1$, $J_2$ and $J_3$ are derived from the direct ab-initio calculation of the 3C, 2H, 4H and 6H polytypes. Indeed in this frame, their respective formation energies per bilayers are:

\begin{equation}
E_{3C} = E_0-J_1-J_2-J_3 
\end{equation}
\begin{equation}
E_{2H} - E_{3C} = 2J_1+2J_3 
\end{equation}
\begin{equation}
E_{4H} - E_{3C} = J_1+2J_2+J_3 
\end{equation}
\begin{equation}
E_{6H} - E_{3C} =\frac{2}{3}J_1+\frac{4}{3}J_2+2J_3 
\end{equation}

Having obtained the parameters $J_1$, $J_2$, and $J_3$, we applied the ANNNI model for the structures 2L, 3L, 4L and 5L. Looking at their tetrahedral sequence as reported in Figure 1 of the main manuscript, we found the relative expressions for the formation energies per unit cell:

\begin{equation}
E_{2L} - E_{3C} = 2J_1+4J_2+6J_3  
\end{equation}
\begin{equation}
E_{3L} - E_{3C} = 4J_1+4J_2+4J_3
\end{equation}
\begin{equation}
E_{4L} - E_{3C} = 6J_1+4J_2+10J_3 
\end{equation}
\begin{equation}
E_{5L} - E_{3C} = 8J_1+4J_2+8J_3
\end{equation}

Results compared with the direct calculation of the supercell energies are reported in Table 1 of the main manuscript.
Notice that the ANNNI model accounts for the interaction between bilayers up to the third nearest neighbour. As a consequence, regarding the evaluation of the formation energy, the 6L structure and beyond (i.e. adding one or more $T'$ tetrahedra to the 5L defect) differ exactly by one or more 2H unit cells. 

\begin{acknowledgement}

The authors acknowledge the CINECA award under the ISCRA initiative for the availability of high-performance computing resources and support.
The authors thank Annalisa Carbone for her valuable assistance with graphs and conducting preliminary calculations.
\end{acknowledgement}

\section{Electronic Supplementary Material}
Supplementary material (additional images and graphs) is available in the online version of this article at http://dx.doi.org/10.1007/***********************).

\bibliography{achemso-demo}

\end{document}